\documentstyle[aps,preprint]{revtex}
\begin{document}
\author{Valery P.Karassiov,}
\address{Lebedev Institute of Physics,
Department of Optics, Leninsky pr. 53, Moscow 117924, Russia, Phone:
(095)1326239, FAX: (095)1357880, e-mail: karas@sci.fian.msk.su}
\author{Vladimir L.Derbov, and Olga M. Priyutova,}
\address
{Department of Physics, Saratov State University, Astrakhanskaya 83, 
Saratov
410071, Russia, Phone: (8452)515195, FAX: (8452)240446, e-mail:
Derbov@scnit.saratov.su, ompri@scnit.saratov.su}
\author{Sergey
I.Vinitsky}
\address{Laboratory of
Theoretical Physics, Joint Institute for Nuclear Research, Dubna,
Moscow region 141980, Russia, Phone:(09621)63348, FAX: (09621)65084,
e-mail: pankova@cvxct0.jinr.dubna.su}   

\title{Polarization coherent states and geometric phases in quantum
optics}
\maketitle
\begin{abstract}

Polarization coherent states (PCS) are considered as generalized
coherent states of $SU(2)_p$ group of the polarization invariance
of the light fields.
The geometric phases of PCS are introduced in a way, analogous to that
used in the classical polarization optics.
\end{abstract}
\newpage

 \section{INTRODUCTION}
\indent
For several recent decades the polarization properties of light have
been widely investigated  both in  theoretical and in applied aspects
(see, e.g., \cite{1,2,3,4,5,6,7,8} and references therein). In 
particular,
some fundamental problems of quantum mechanics, related to the ``hidden''
variables, Bell's inequalities and Einstein-Podolsky-Rosen paradox,
quantum chaos, Berry's and other geometric phases, etc., are successfully
studied by means of quantum polarization optics.

It is well known \cite{3,9,10} that the generalized coherent states 
(GCS),
generated by the action of the displacement operators $D(g)=\exp (
\sum d_iF_i)$ of the groups $G^{DS}$ on certain fixed reference vectors
$|\psi_0\rangle$ in the given space $L^D$ of the representation 
$D(G^{DS})=
\{ D(g), g\in G^{DS} \}$ of  the groups $G^{DS}$, present an effective
tool for the study of quantum systems having the dynamic symmetry (DS)
groups $G^{DS}$. In particular, the average values $\langle\{\alpha_i\};
\psi_0|f(\{ F_i\})|\{\alpha_i\};\psi_0\rangle$ of the arbitrary
functions $F(\{ F_i\})$,
corresponding to the observables and depending on the generators
$F_i$ of $G^{DS}$, as well as the quasiprobability distribution
functions
$Q(\{\alpha_i\};\psi_0)_\rho=\langle\{\alpha_i\};\psi_0|\rho|
\{\alpha_i\};\psi_0\rangle$,
$\rho$ being the density matrix, are widely used for the
description of quasiclassical properties of the  appropriate
quantum systems near the ``classical limit'' \cite{2,10}. For
example, in quantum optics similar quantities, defined using the
conventional Glauber's coherent states and associated with
Weyl-Heisenberg group $W(m)$, are widely  used for the description
of $m$-mode electromagnetic fields \cite{2,3}. The GCS associated
with $SU(m)$ groups play the same role for the
systems of $n$-level emitters of radiation \cite{3,9}.

Recently it was shown \cite{1,3,7} that the DS group adequate to
the polarization properties of quantum light is the $SU(2)_p$
group of the polarization invariance of the free light fields.
It's generators $P_\alpha,
\quad \alpha=0,1,2$ (or $\alpha=0,\pm$) are
the components of the polarization (P) quasispin, which corresponds to
the Stokes vector $\vec{\Sigma}=(\Sigma_{\alpha})$ \cite{6}, 
parameterized
on the so-called Poincare sphere $S_{P}^2$
\cite{4,8} in the classical statistical optics.

The aim of the present paper is to  investigate the GCS
of the $SU(2)_p$ group (see also \cite{7,11}) in the $2m$-mode Fock
space
$L_F(2m)=\mbox{Span}\{\prod_{j=1}^m(a_+^+(j))^{n^+_j}(a^+_-(j))^{n^-_j}
|0\rangle\}$
with two polarization and $m$ spatiotemporal (ST) modes in the helicity
$(\pm)$ polarization basis in both mathematical and physical aspects.
An  attention is paid to differences between pictures of independent
(uncorrelated) and correlated ST modes which correspond, respectively,
to one-mode and broad-band measuring devices (cf. \cite{12}); from the
mathematical viewpoint these cases differ by using collective 
($P_{\alpha}=
\sum_{j=1}^m P_{\alpha}(j), S^2_P$) or "individual" ($P_{\alpha}(j),
S^2_{P}(j)$) P-quasispin components and Poincare spheres.
Then Pancharatnam-type geometric phase acquired
by these states during the cyclic evolution on the Poincar\'e sphere
is derived and compared with the classical results \cite{8}.
We also briefly discuss some other applications of polarization GCS to
the quasiclassical description of the polarization properties of
quantum light.

\section{POLARIZATION COHERENT STATES}

The general definition of GCS of the $SU(2)$ group is well known
\cite{9,10}
\begin{eqnarray}
|\xi ;\psi_0\rangle \equiv |\theta,\varphi;\psi_0\rangle=\exp (\xi J_+ -
\xi^* J_-)|\psi_0\rangle,
\label{e1}
\end{eqnarray}
where $\xi=-\theta /2 \exp(\it -i\varphi), \quad 0\le \theta \le\pi,
\quad 0\le \varphi \le 2\pi$ are the angular coordinates of the
``classical'' quasispin $\vec J=(J_\alpha)$ in its ``phase''
space, i.e. the Poincare sphere $S^2_{P}(\theta,\varphi)$;
$|\psi_0\rangle$ is a certain reference
vector in the space $L$ of the states of the system. From the
physical viewpoint, the states $|\xi ;\psi_0\rangle$ describe
output light beams obtained by means of action of quantum
"$SU(2)$-rotators" with Hamiltonians

\begin{eqnarray}
H_{SU(2)}= g J_+ +g^* J_-
\label{e2}
\end{eqnarray}
on the input beams in the quantum state $|\psi_0\rangle$ (see, e.g.,
\cite{5} and references therein for possible realizations of such
rotators in experimental devices for $m=1$).

For spin systems having a fixed spin value $j$ one of the basis
vectors $|jm\rangle$ of the irrep $D^j(SU(2))$ is used as
$|\psi_0\rangle$, and the values of $m=\pm j$ correspond to the
``squeezed'' GCS most near to classical states \cite{10}.
A peculiarity of the polarization spin $(J_\alpha = P_\alpha)$ of
the light fields is that the Fock spaces $L_F(2m)$ of the field
states may be viewed as direct sums of the specific $SU(2)$ fiber
bundles and contain the subspaces $L^{j\sigma}$ of the irrep
$D^j(SU(2))$ with $j=p=0,1/2,1,$\dots , generally (in the picture
of correlated ST modes) with a certain multiplicity, where the index
$\sigma $ labels the $SU(2)$-equivalent subspaces $L^{j,\sigma}$
and corresponds to the additional (non-polarizational) degrees
of freedom \cite{1,7}. Hence to get the ``complete phase
portrait`` of the quantum light field in all $L_F(2m)$ one should
have complete sets of the GCS of $SU(2)_p$ similar to Eq.~(\ref{e1}) with
a set of reference vectors
$|\psi_0\rangle=|\psi_0^{p,\sigma}\rangle\in L^{p,\sigma},
\quad p=0,1/2,\dots$, or with a set of vectors
$|\psi_0\rangle=|\psi_{0,\gamma}\rangle$, having
nonzero  projections on each of the subspaces $L^{p,\sigma},\quad
p=0,1/2,...$.

In the first case, basing on the maximal classicality criterion
in the polarization degree of freedom it seems natural to choose
(in the picture of correlated ST modes) for $|\psi_0^{p,\sigma}\rangle$
the vectors

\begin{eqnarray}
|p,\pi  =  \pm p;n,\lambda\rangle
 = \sum_{\sum \alpha^\pm_j=2p,\sum
\gamma_{ij}=\frac{n}{2}-p} C(\{\alpha^\pm_j; \gamma_{ij}\})
\prod^m_{j=1}(a^+_\pm(j))^{\alpha^\pm_j} \prod_{i<j}
(X^+_{ij})^{\gamma_{ij}}|0\rangle,
\label{e3a}
\end{eqnarray}
where $X^+_{ij} \equiv a^+_+(i)a^+_-(j)-a^+_-(i)a^+_+(j)$ are the 
creation
operators of the $SU(2)$-invariant two-photon clusters.
These vectors belong to a polarization-invariance adapted
orthonormalized basis in $L_F(2m)\quad \{|p,\pi;n,\lambda\rangle\}$, 
which
is defined by the following equations \cite{1,7,12}
\begin{eqnarray}
P^2|p,\pi;n,\lambda\rangle & = & 
p(p+1)|p,\pi;n,\lambda\rangle,\nonumber\\
P_0|p,\pi;n,\lambda\rangle & = & \pi|p,\pi;n,\lambda\rangle, 
\label{e3b}\\
 N|p,\pi;n,\lambda\rangle & = & n|p,\pi;n,\lambda\rangle,\nonumber
\end{eqnarray}
where $P^2={1\over 2}(P_+P_-+P_-P_+)+P^2_0$ is the $SU(2)$ Casimir
operator, $P_\pm =\sum^m_{j=1}P_\pm(j)=\sum^m_{j=1}a^+_\pm(j)a_\mp(j),
P_0 =\sum^m_{j=1}P_{0}(j)={1\over2}\sum^m_{j=1}[a^+_{+}(j)a_{+}(j)-
a^+_{-}(j)a_{-}(j)]$,  $N=\sum^m_{j=1}[a^+_{+}(j)a_{+}(j)+
a^+_{-}(j)a_{-}(j)]$ is the total photon number, and $\lambda$ is the
extra quantum label. In particular cases $m=1,2$ the vectors given
by Eq.~(\ref{e3a}) take the form \cite{7,11}:
\begin{eqnarray}
|p\rangle_\pm \equiv
|p,\pi=\pm p; n=2p\rangle=[(2p)!]^{-{1\over 2}}(a^+_\pm
(1))^{2p}|0\rangle,\quad 2p=0,1,\dots, \label{e4}
\end{eqnarray}
and
\begin{eqnarray}
&&|p,n,t\rangle_\pm  \equiv   |p,\pi=\pm p;n,t\rangle \nonumber \\
&& =  \left[{(n/2+p+1)!(n/2-p)!(p+t)!(p-t)! \over
(2p+1)!}\right]^{-{1\over 2}}
(a^+_\pm(1))^{p+t}(a^+_\pm(2))^{p-t}(X^+_{12})^{n/2-p}|0\rangle,
\label{e5}
\end{eqnarray}
where $2t$ is the difference $N(1)-N(2)=N_+(1)+N_-(1)-N_+(2)-N_-(2)$
of the photon numbers in the first and second ST modes.

Now making use of the definition given by Eq.~(\ref{e1})
and of the transformation
properties of the operators $a^+_\pm(j), X^+_{ij}$ with
respect to the group $SU(2)_p$ \cite{1,7} we get the sets of the
polarization GCS generated by the reference vectors (\ref{e3a}) (compare
with \cite{7}):
\begin{eqnarray}
|\theta,\varphi;p,n,\lambda \rangle_\pm  & \equiv  & \exp(\xi 
P_+-\xi^*P_-)
|p,\pm p;n,\lambda\rangle  \nonumber \\
& = & \sum C(\{\alpha^\pm _j,\gamma_{ij}\})
\prod^m_{j=1} (a^+_{\pm}(j;\theta,\varphi))^{\alpha^\pm_j}
\prod_{i<j} (X^+_{ij})^{\gamma_{ij}}|0\rangle, \label{e6}
\end{eqnarray}
where  the operator $a^+_{\pm}(j;\theta,\varphi) \equiv
a^+_\pm(j)\cos {\theta\over 2} \pm a^+_\mp(j) \exp(\pm i\varphi)
\sin{\theta\over 2}$ may be interpreted as the creation operator of
the elliptically polarized photon in the $j$-th ST mode having
the ellipticity parameters defined by the angles $\theta,\varphi$ 
\cite{6}.

Since the quasispin $\vec P$ is the vector of SU(2) space one can see
from the analogy with the theory of transformation of angular momentum 
that from the mathematical point of view the action of
polarization rotator is equivalent to the multiplication of the 
initial state vector by the spherical function of the finite 
rotation of the first order.  

Changing the direction averaged over the quasispin state  will be
analogous to turning the coordinate system by $\theta$ and $\varphi$
angles and can be expressed in the following way
\begin{eqnarray} 
\langle \xi, \psi_0 | P_{\alpha} |\xi, \psi_0 \rangle
= \sum_{\beta} D^1_{\alpha \beta} (\theta, \varphi) \langle
\psi|P_{\beta} | \psi \rangle , \label{40} 
\end{eqnarray} 
where $D^1_{\alpha \beta}(\theta, \phi)$ is the Wigner function.

As follows from this expression, according to the group theory,   
the sum of the squares of the mean values of the  polarization
quasispin components will be $SU(2)$-invariant: 
 \begin{eqnarray}
  \sum_{\alpha}\langle \xi ,
\psi_0| P_{\alpha}| \xi, \psi_0\rangle^2 ={\it C}= R^2, \label{41}
\end{eqnarray}
 where the constant $R=\sqrt{ \sum_\alpha \langle \psi_0|P_\alpha|\psi_0
 \rangle ^2}$ is the radius of the polarization sphere (Poincar\'e).

 Indeed, averaging the polarization quasispin components over
 states ($\ref{e4}$)  
 leads to the conventional parameterization of Poincar\'e sphere
 (when the poles correspond to the states with circular polarization):
 $$\langle \xi;\psi_0 | P_0 | \xi ;\psi_0 \rangle = R\cos{\theta},$$
 \begin{eqnarray}
 \langle \xi; \psi_0 | P_1 | \xi ;\psi_0 \rangle =
 R\sin{\theta}\cos{\phi}, \label{e80}
 \end{eqnarray}
 $$ \langle \xi; \psi_0 | P_2 | \xi ;\psi_0
 \rangle = R\sin{\theta}\sin{\phi};$$
 where the parameters $\theta$ and $\varphi$ are the angular coordinates 
of
 the point on the sphere with the radius 
$R=\langle \psi_0 |P_0|\psi_0 \rangle$.
 This point defines the direction of vector of "classical" quasispin
 after the single ST mode with minimum value of polarization uncertainty
 have passed  through the polarization rotator.

 The action of this rotator on the state of this type with $m=2$
  may be described in the follows way
\begin{eqnarray}
|\theta,\varphi;p,n,t\rangle_{\pm}\equiv|\theta,\varphi;p,\pi=\pm 
p;n,t\rangle
=\left[\frac{(n/2+p+1)!(n/2-p)!(p+t)!(p-t)!}{(2p+1)!}\right]^{-1/2}\nonumber\\
\times(a_{\pm}^+(1;\theta,\varphi))^{p+t}
(a_{\pm}^+(2;\theta,\varphi))^{p-t}
(X^{+}_{12})^{n/2-p}|0\rangle,
\label{e100}
\end{eqnarray}

The sets of the polarization GCS given by Eq.~(\ref{e6}) belong , from
the mathematical point of view, to the class of semi-coherent ones
( which are coherent (quasiclassical) in polarization degrees of freedom
and orthonormalized (strongly quantum) in other ones) \cite{7}, and
yield the following decomposition of the identity operator $\hat I$
in $L_F(2m)$ \cite{5,7}, which is an expression of the basis set
completeness:
\begin{eqnarray}
\hat I = \sum_{n,p,\lambda} \int_0^\pi \int_0^{2\pi}{(2p+1)
\over 4\pi} \sin \theta d\theta d\varphi
|\theta,\varphi;p,n,\lambda\rangle_\pm\langle\theta,\varphi;p,n,\lambda|_\pm.
\label{e7}
\end{eqnarray}
Choosing for $|\psi_0^{p,\sigma}\rangle$ the reference vectors different
 from those given by Eq.~(\ref{e3a}), one may construct, by means of
Eq.~(\ref{e1}), other sets of GCS of the $SU(2)_p$ group, which in the
mathematical aspect are equivalent to Eq.~(\ref{e6}) \cite{7,11}.

The construction of Eq.~(\ref{e6}) is simplified in the picture of
independent ST modes, when the group $SU(2)_p$ acts in the space
$L_F(2)$ of each $j$-th ST mode independently, and its action is
determined by the angles $(\theta_j,\varphi_j)$;
\begin{eqnarray}
|\{\theta_j,\varphi_j\};\{n_j\}\rangle_\pm  & \equiv  & \prod^m_{j=1}\exp
(\xi_jP_+(j)-\xi^*_jP_-(j))(a^+_\pm(j))^{n_j}|0\rangle \nonumber \\
 & = & \prod_{j=1}^m
{(a^+_{\pm}(j;\theta_j,\varphi_j))^{n_j}\over (n_j!)^{1/2}}|0\rangle.
\label{e8}
\end{eqnarray}
The set of GCS (\ref{e8}) is complete (an analog of Eq.~(\ref{e7})
is valid for it) and yields the ``polarization phase portrait of the
field'' adequate to independent measurements for each ST mode. The
connection between the sets (\ref{e6}) and (\ref{e8}) is realized via the
generalized Clebsh-Gordan coefficients of $SU(2)_p$ \cite{9}.

We note that states (\ref{e8}) are the Fock states in terms of
"rotated" photon operators $a^+_{\pm}(j,\theta,\varphi)$ and are
unitarily equivalent to initial states (\ref{e4}); hence  there are
some difficulties to produce them (as well as states (\ref{e6}) in
physical experiments. Therefore, from the physical viewpoint it is of
interest  to consider alternate types of GCS of $SU(2)_p$, which do not
contain the discrete parameters $n,\lambda$. For example, if instead
of Eq.~(\ref{e3a}) one takes the vectors $|\psi_0^{p,\{z\}}\rangle=
\exp(\sum_i z_i F_i)(a^+_\pm(1))^{2p}[(2p)!]^{-1/2}|0\rangle$,
$F_i$ being the generators of the $SO^*(2m)$ group, complementary
to $SU(2)_p$), we obtain states which are  GCS with respect to both
$SU(2)_p$ and $SO^*(2m)$ groups and display a specific squeezing
in both polarization and biphoton degrees of freedom \cite{12}.

Another type of such "physical" polarization GCS may be obtained if
instead of Eq.~(\ref{e3a}) one takes the sets of reference vectors
$|\psi_{0,\gamma}\rangle$, having the nonzero projections on all
$L^{p,\sigma}$. An example of such a set is the familiar set of
Glauber's coherent states:
\begin{eqnarray}
|\{\alpha^+_j,\alpha^-_j\}\rangle=\prod^m_{j=1} \exp
[\alpha^+_ja^+_+(j)+\alpha^-_ja^+_-(j)-
(\alpha^+_j)^*a_+(j)-(\alpha^-_j)^*a_-(j)]|0\rangle.
\label{e9}
\end{eqnarray}
Then using the definition (\ref{e1}) and taking the $SU(2)_p$
transformation properties of $a^+_\pm(j)$ into account, one gets
from Eq.~(\ref{e9}) the set of GCS
\begin{eqnarray}
|\theta,\varphi;\{\alpha_j^+,\alpha_j^-\}\rangle
 & \equiv & \exp(\xi P_+-\xi^* P_-)
|\{\alpha^+_j,\alpha^-_j\}\rangle=|\{\tilde \alpha^+_j(\theta,\varphi),
\tilde \alpha^-_j(\theta,\varphi)\}\rangle, \label{e10}        \\
\tilde \alpha^\pm_j(\theta,\varphi) & = & 
\alpha^\pm_j\cos\frac{\theta}{2}
\mp\exp(\mp i\varphi)\alpha^\mp_j\sin\frac{\theta}{2}, \nonumber
\end{eqnarray}
which can be obtained experimentally by the action of quantum 
polarization
"rotators" described by Hamiltonians of the form
\begin{eqnarray}
H_{SU(2)_p} &=& \sum_{\alpha, \beta =
\pm}\zeta_{\alpha,\beta} \sum_{j=1}^{m} a^+_{\alpha}(j) a_{\beta}(j)
\label{e2^*}
\end{eqnarray}
on the initial states (\ref{e9}).

The states (\ref{e10}) are analogous to the initial set (\ref{e9}),
but with $SU(2)$ -rotated parameters $\alpha^{\pm}_{j}$ involving two
additional (redundant from the mathematical viewpoint)
parameters $\theta$ and $\varphi$ that is of no importance from the
physical viewpoint. We also note that states (\ref{e10}) can be 
represented
in the form:
\begin{eqnarray}
|\{\alpha^+_j,\alpha^-_j\}\rangle=\prod_{j=1}^m |\theta_j,\varphi_j;
\alpha_j\rangle_+,
\nonumber\\
|\theta_j,\varphi_j;\alpha_j\rangle_+ \equiv
|\theta_j,\varphi_j;\alpha^+_j,0\rangle&=& 
|\alpha^+_j\cos\frac{\theta_j}{2},
\alpha^+_j\exp(i\varphi_j)\sin\frac{\theta_j}{2}\rangle,\label{e11}
\end{eqnarray}
where the polarization coordinates are picked out explicitly (unlike in
the form (\ref{e10}))


\section{GEOMETRIC PHASES OF POLARIZATION
COHERENT STATES}

In classical polarization optics it is well known \cite{8,13}, that
during the cyclic evolution of its polarization state the
classical plane wave acquires an additional phase shift equal to
half the solid angle subtended by the trajectory of the tip of the
Stokes vector on the Poincar\'e sphere. This additional phase is
shown to be a particular case of the Pancharatnam's phase \cite{14}
associated with the $SU(2)$ symmetry of the polarization states.
It is invariant with respect to deformations of the trajectory
leaving the solid angle unchanged and, therefore, is of purely
geometric nature. Pancharatnam's ideas have been used  \cite{15}
to set a generalized definition of the geometric phase, valid for a
wide class of quantum evolutions, generally, non-cyclic.

It is natural to pose a question, what happens to the states of
quantum light in the similar situation. The considerations
presented above make it possible to apply the general
definitions \cite{15,16} of the geometric phase $\gamma$ to the 
polarization
GCS, since the angles $\theta, \varphi$ enter the appropriate
expressions explicitly as classical parameters:
\begin{eqnarray}
\gamma=-\oint_C A_sds, \label{e13}
\end{eqnarray}
where the gauge potential $A_s$ is expressed as
\begin{eqnarray}
A_s & = & \mbox{Im}\langle\xi(\theta,\varphi),\psi_0
|\frac{d}{ds}|\xi(\theta,\varphi),\psi_0\rangle \nonumber \\
 & = &
\mbox{Im}\langle\xi(\theta,\varphi),\psi_0|\nabla_
{\vec\Omega}|\xi(\theta,\varphi),\psi_0\rangle
{d\vec \Omega\over ds},
\label{e14}
\end{eqnarray}
$s$ is an evolution variable which determines the motion of the system
along the evolution trajectory $C$. The states
$|\xi(\theta,\varphi),\psi_0\rangle$
are supposed to be the normalized polarization GCS defined
by Eq.~(\ref{e1}) with a certain
particular choice of the reference state vectors mentioned above.
The explicit form of the derivatives on the unit sphere may be written as
\begin{eqnarray}
{d\vec \Omega\over ds}=
\vec e_\theta\frac{d\theta}{ds}+\vec e_\varphi
\sin\theta\frac{d\varphi}{ds},\label{e15}
\end{eqnarray}
and
\begin{eqnarray}
\langle\xi(\theta,\varphi),\psi_0|\nabla_{\vec\Omega}|\xi(\theta,
\varphi),\psi_0\rangle
=
\left[\frac{\vec e_\theta}{\theta}\frac{\partial}{\partial u}+
\frac{\vec e_\varphi}{\sin\theta\varphi}
\frac{\partial}{\partial v}\right]
\langle\xi(\theta,\varphi),\psi_0|\xi(u\theta,v\varphi),\psi_0\rangle
_{u=v=1}.
\label{e16}
\end{eqnarray}

As the first example let us consider the polarization GCS (\ref{e6})
with $m=1,2$ and the reference state vectors given by Eqs.~(\ref{e4}),
(\ref{e5}). The overlap integral in Eq.~(\ref{e16}) then takes the form
\begin{eqnarray}
\langle\xi(\theta,\varphi),\psi_0|\xi(u\theta,v\varphi),\psi_0\rangle
& \equiv &
\langle\theta,\varphi;p,2p|u\theta,v\varphi;p,2p\rangle_{\pm}\nonumber \\
& = &\left[\cos\frac{\theta}{2}\cos\frac{u\theta}{2}+\sin\frac{\theta}{2}
\sin\frac{u\theta}{2}
\exp(\mp i \varphi(v-1)\right]^{2p}.
\label{e17}
\end{eqnarray}
where $p=n/2$ for $m=1$ and $0\leq p \leq n/2$ for $m=2$.
The derivatives are easily calculated explicitly for any $p$, and from
Eqs.~(\ref{e13})-(\ref{e17})  we obtain
\begin{eqnarray}
\gamma=\pm 2p \oint_C\sin^2\frac{\theta}{2}d\varphi, \label{e18}
\end{eqnarray}
which is the $2p$ multiple to half the solid angle subtended by $C$ on
the Poincare sphere. In particular, for $p=1/2$ this result coincides 
with that
for  classical plane waves.

For the GCS given by Eq.~(\ref{e10}) with the reference state vectors
(\ref{e9}) one gets
\begin{eqnarray}
&&\langle\xi(\theta,\varphi),\psi_0
|\xi(u\theta,v\varphi),\psi_0\rangle\equiv
\langle\theta,\varphi;\{\alpha_j^+,\alpha_j^-\}|
u\theta,v\varphi;\{\alpha_j^+,\alpha_j^-\}\rangle \nonumber \\
&&=\exp \left[-\frac{1}{2}\sum_{j=1}^m
\{|\tilde\alpha_j^+(\theta,\varphi)|^2
+|\tilde\alpha_j^-(\theta,\varphi)|^2
+|\tilde\alpha_j^+(u\theta,v\varphi)|^2
+|\tilde\alpha_j^-(u\theta,v\varphi)|^2 \} \right] \nonumber \\
&&\times\exp\left[\sum_{j=1}^m\{\tilde\alpha_j^+(\theta,\varphi)
(\tilde\alpha_j^+(u\theta,v\varphi))^*+
\tilde\alpha_j^-(\theta,\varphi)
(\tilde\alpha_j^-(u\theta,v\varphi))^*\}\right],\label{e19}
\end{eqnarray}
where the notation $\tilde\alpha_j^\pm(\theta,\varphi)$ is clear
from Eq.~(\ref{e10}). Making use of
Eqs.~(\ref{e14})-(\ref{e16}),(\ref{e19}) we get the explicit
expression of the gauge potential $A_s$
\begin{eqnarray}
A_s & = &  A_s^{(1)}+A_s^{(2)};\label{e20} \\
A_s^{(1)} & = & -\frac{d\varphi}{ds}\sin^2\frac{\theta}{2}
\sum_{j=1}^{m}\left(|\alpha_{j}^{+}|^2-|\alpha_{j}^{-}|^2\right);
\label{e21}\\
A_s^{(2)} & = & -Re\left[\left(\sin\theta\frac{d\varphi}{ds}
+i\frac{d\theta}{ds}\right)exp(-i\varphi)
\sum_{j=1}^{m}\alpha_{j}^{-}(\alpha_{j}^{+})^*\right].
\label{e22}
\end{eqnarray}

When going from presentation in circular basis  
to Cartesian basis for polarization quasispin component the final
 expression for geometric phase takes form
\begin{eqnarray}
\gamma = \gamma^{(0)} +\gamma^{(1)} + \gamma^{(2)}  \label{22} \\
\gamma^{(0)}=2\langle P_0\rangle \oint_C
\sin^2\frac{\theta}{2}d\varphi,  \quad
2\langle P_0\rangle = \sum_{j=1}^{m}
\left(|\alpha_{j}^{+}|^2-|\alpha_{j}^{-}|^2\right),
\label{23} \\
\gamma^{(1)}= - \langle P_1\rangle \oint_C
  [\sin \theta \cos\varphi d\varphi +
\sin \varphi d\theta],\quad
\langle P_1\rangle = Re [\sum_{j=1}^{m}\alpha_{j}^{-}(\alpha_{j}^{+})^*],
\label{24} \\
\gamma^{(2)}= \langle P_2\rangle \oint_C  [\sin\theta \sin\varphi 
d\varphi -
\cos\varphi d\theta], \quad
\langle P_2\rangle = -Im [\sum_{j=1}^{m}\alpha_{j}^{-}(\alpha_{j}^{+})^*]
\label{25}
\end{eqnarray}

Therefore the geometric phase is  expressed 
as a scalar product of two vectors. The first one is the vector of 
polarization quasispin averaged over initial Glauber state (\ref{e9}).
This vector corresponds to the quantum polarization state of the input 
beam
and represents the possible nonclassical properties of light source. 
The second vector consists of contour integrals on   
Poincar\'e sphere, does not depend on the state of the light beam and 
characterizes transformation of polarization on the track.

In particular, for PGCS (\ref{e3a}) with $p=n/2$ and $m=1,2$  we have 
 $\langle P_{0}\rangle=p,\quad \langle P_{1}\rangle=0,\quad 
\langle P_{2}\rangle=0        $, that leads 
to (\ref{e18}).

It may be seen that the contribution of $A_{s}^{(1)}$ to the geometric
phase (\ref{e13})
is just the classical half the solid angle subtended by the cyclic
evolution loop $C$  on the Poincar\'e  sphere, multiplied
by a factor depending on  mode structure of the
field. If for each $j$ either $\alpha_{j}^{+}$ or $\alpha_{j}^{-}$
equals zero then $A_{s}^{(2)}$ vanishes, and Eq.(\ref{23}) represents
 the total geometric phase. This is valid, in particular,
for the single-mode states (\ref{e11}) associated to elliptically
polarized waves obtained after transmissions of coherent light beams with
a definite circular polarization ($<|P_{0}|> \not= 0 ,<|P_{1}|> = 0 ,
<|P_{2}|> = 0 $) through polarization rotators.
At the same time it is not the case for general GCS (\ref{e10}) with
initial (reference) vectors (\ref{e9}) ;
specifically, even in the case of one ST mode
$A_{s}^{(2)}$ does not vanish that reflects a specific correlation
of polarization modes after a transmission of beams (\ref {e9})
with $\alpha^{\pm}_j$ being purely real through "polarization
rotators" (\ref{e2^*}). A similar situation
(when $A_{s}^{(2)}$ does not vanish) also occurs in the case
$|\alpha_{j}^{+}| = |\alpha_{j}^{-}|$ in Eqs. (\ref{e9}),(\ref{e10})
corresponding to a transmission of "helicityless" ( $<|P_{0}|> = 0 $)
 coherent light beams \cite {1,12} through polarization
rotators. In general, Eqs. (\ref{e20}-\ref{e22})
describe a structure (nature) of influences of polarization
rotators  on initial Glauber's CS in dependence on their polarization
properties since "energetic" multipliers in these equations are
related to expectation values of different components of
polarization quasispin $P_{\alpha}$.\\

\section{DISCUSSIONS AND IMPLICATIONS}

The sets of GCS obtained above may be also used for the quasiclassical
analysis of the polarization properties of quantum light fields. In
particular, following the general rules \cite{2,10} one can use the
definition (\ref{e1}) to introduce the complete polarization distribution
functions of the quasiprobability \cite{11} as
\begin{eqnarray}
Q(\theta,\varphi;\psi_0)_\rho\equiv\langle\theta,\varphi;\psi_0|\rho|
\theta,\varphi;\psi_0\rangle\equiv
\mbox{Tr}[\rho |\theta,\varphi;\psi_0\rangle\langle
\theta,\varphi;\psi_0|], \label{e24}
\end{eqnarray}
where $\rho$ is the complete density matrix for the state of the
field, $|\theta,\varphi;\psi_0\rangle$ being defined by Eq.~(\ref{e1}). 
Then,
substituting the specifications (\ref{e6}), (\ref{e8}),
(\ref{e10}), (\ref{e11}) for
$|\theta,\varphi;\psi_0\rangle$ found above into Eq.~(\ref{e24}), we get
the appropriate concrete  types of the complete polarization
quasiprobability functions. Note, however,  that such functions,
besides the dependence on the polarization parameters $\theta,
\varphi$, involve the additional quantum numbers $n,p,\lambda,
\{\alpha^\pm_j\}$, etc., which characterize the non-polarization
degrees of freedom of the field. Therefore, to obtain its ``polarization
quasiclassical portrait'' in the $\rho$-state one may make use of the
reduced polarization quasiprobability functions $Q(\varphi,\theta;\psi_0)
^r_\rho$, resulting from  Eq.~(\ref{e9}) after the summation (or 
integration)
over the non-polarization variables. Such functions may be used to
analyse the ``polarization squeezing'' \cite{12} in analogy with the 
familiar
$Q$-functions in case of the standard quadrature squeezing \cite{17}.

So, we have demonstrated the presence of geometric phase in the
polarization generalized coherent states. This phase is due to
the cyclic evolution in the space of the parameters
$\theta, \varphi$, which are the angular coordinates of the
``classical'' quasispin on the Poincare sphere. The explicit
expression of the geometric phase is shown to depend on the
polarization quasispin $p$ and the choice of the reference state
vectors, and only in the particular case of $p=1/2$ , the GCS generated 
by
the set of essentially non-classical reference vectors (\ref{e3a})
demonstrate exactly the same geometric phase as in
the classical case studied by other authors \cite{13}.
For particular sets of GCS generated by Glauber's CS
the geometric phase is shown to be a product
of the classical expression by a factor depending on the mode structure
in the picture of independent ST modes, whereas the result is  different
for GCS generated by multimode Glauber's CS in the picture of
correlated ST modes.

Thus we have presented a straightforward generalization of the
classical theory of the geometric phase induced by the
evolution of the polarization onto a new class of
quantum light states.
The expressions derived are to be used in further investigation
of the geometric phases in quantum optics.  Specifically, it is
of interest to calculate geometric phases for different types
of non-classical states of unpolarized and partially polarized
light displaying polarization squeezing \cite{12} that may
prove to be useful for a practical identification of these states.

We also note that the difference of Eqs. (\ref{e20})-(\ref{e22})
from the classical results and Eq. (\ref{e18})
reflects the fact the states (\ref{e9}) and (\ref{e10}) are
essentially less quasiclassical
as compared to GCS (\ref{e6}) {\it with respect to polarization degrees
of freedom}. Therefore Glauber's CS do not simulate completely classical
light waves.
 
To compare our results correctly with those known in classical 
polarization 
optics
\cite{13} it is
necessary to emphasize that  the phase measured in usual interference 
experiments
is not the phase of the field state vector but the phase acquired 
by the field amplitude operator of the field state vector \cite{30}. 
As follows from numerous examples from \cite{30}, the latter may
be identified with the classical Hannay angle \cite{31}. The
relation between the Hannay angle and the Berry phase of the 
corresponding
quantum states has been considered in several papers 
\cite{32,33,34}. The rigorous relation follows from
the decomposition of the Berry's phase in powers of $\hbar$ in the
quasiclassical limit. 
If one omits the terms of the order $\hbar^2$ and higher, the Hannay
angle  $h_j$, associated with the $j$-th classical degree of freedom,
appears to be equal to
\begin{eqnarray}
 h_j = -\frac{\partial\gamma_{n_j}}{\partial n_j},  \label{26}
\end{eqnarray}
where $\gamma_{\{n_j\}}$ is the geometric phase of the state
having the set of quantum numbers $\{n_j\}$ which are coupled to
the corresponding classical action variables via the Bohr-Zommerfeld
condition $I_j=(n_j+\mu_j)\hbar$  \cite{34}. 

In our case $I_j$ is a classic integral of motion in the space defined 
by the averaged components of polarization quasispin $\langle
P_{\alpha}\rangle (\alpha=0,1,2) $, {\it i.e.} $I_p=\hbar R $,
where $R=\sqrt{\langle P_0 \rangle^2+ \langle P_1 \rangle^2+\langle P_2
\rangle^2}$ is the radius of Poincar\'e sphere.
Correspondingly, the expression for polarization  Hannay angle, according 
with  (\ref{e80}), (\ref{23})-(\ref{25}) and (\ref{26}), have the form
\begin{eqnarray}
h_p=2\cos\theta_0\oint_C\sin^2\frac{\theta}{2}d\varphi-
\sin\theta_0\cos\varphi_0\oint_C(\sin\theta\cos\varphi d\varphi +
\sin\varphi d\theta) \nonumber \\
+\sin\theta_0\sin\varphi_0\oint_c(\sin\theta\sin
\varphi d\varphi-\cos\varphi d\theta), \label{e40}
\end{eqnarray}
where $(\theta_0,\varphi_0)$ are the angular coordinates of the
initial position of the classical polarization quasispin vector
on the Poincar\'e sphere.

  Now it becomes clear that the geometric phase of the
''polarizationally most classical'' PGCS  (\ref{e18}) completely agrees
with the classical results
\cite{13} since after
the calculation of the derivative (\ref{26}) in accordance with
\cite{30} it follows from  (\ref{e18}) that the phase shift
observed in a usual interferometer (Hannay angle) is just half
the solid angle on the Poincar\'e sphere.

   For the PGCS obtained from the Glauber states the connection
between the geometric phase (\ref{22})-(\ref{25}) and the phase
observed in a usual interferometer is less evident, since the
state-dependent factors multiplied by the contour integrals
on the Poincar\'e sphere are no longer ``good'' quantum numbers, and
the states themselves are strongly nonclassical in the polarization
degrees of freedom. However, it may be shown that the phase acquired
by the field amplitude (Hannay angle) will be half the solid angle
subtended by $C$ again.

  Therefore, the quantum nature of light doesn't manifest itself
in the experiments like  \cite{8} where the
total intensity of the light is measured at the output of the
interferometer even if one proceeds to the photon-counting technique.
This result agrees with the conventional point of view
\cite{5}.

However, new phase information may be obtained if
at the output of the interferometer one measures physical
observables other than the total intensity, for example,
the components of the P-quasispin or the polarization
noise. Even in the simplest case of Fock states the result is
different from the usual one. Indeed, according to  \cite{30}
during the cyclic evolution the field operators acquire the phase
factors
$a^{+}_{+}\rightarrow \tilde a^{+}_{+}=e^{ih_+}a^{+}_{+},\quad
a^{+}_{-}\rightarrow \tilde a^{+}_{-}=e^{ih_-}a^{+}_{-}$ with
different Hannay angles for the right- and left-hand-polarized
components. At the output the superposition of the fields is
described by the operators
$\hat a^{+}_{\pm}=ra^{+}_{\pm}+t\tilde a^{+}_{\pm}$.
Then both  $h_{\pm}$ themselves and their combinations
$h_{+}\pm h_{-}$ appear in the mean values of $ P_{1,2}^{k} $
composed of these operators  using the general definition.

The search for the ways to measure the GPs of the state
vectors of quantum light, particularly, PGCS,
as well as for most simple and convenient technologies of the
experimental realization of multimode
quantum polarization rotators is a subject of our further studies.
One of the promising approaches is suggested in \cite{35}.
This approach involves a combination of
experimental two-photon interferometer setups, aimed at the study
of entangled states, with the ideas of the geometric phases
defined according to \cite{15}.

In conclusion we note that the polarization GCS of $SU(2)_p$ group
obtained may be applied also to the analysis
of other aspects of the polarization quasiclassical description.
Among them one should mention the polarization uncertainty relations
and so-called intelligent states \cite{18}, as well as the
general problem of the
description of the quantum light field phase \cite{19}.

\section{ACKNOWLEDGMENTS}

We are grateful to Dr. A.V.Masalov, Dr. K.I.Guriev 
and  Dr.A.V.Gorokhov for helpful discussions.

This work was partially supported by the State Committee for High School 
of
Russia, grant No. 95-0-2.1-59.


\begin{references}
\bibitem{1}
V.P.Karassiov,
Polarization structure of quantum
light: a new insight. 1. General outlook. Preprint FIAN No.63,
(Moscow, 1992);  J.Phys. A 26 (1993) 4345.
\bibitem{2}
J.Perina,  Quantum statistics of linear and
nonlinear optical phenomena (Reidel, Dardrecht, 1984).
\bibitem{3}
V.P.Karassiov and V.I.Puzyrevsky,
 Trudy FIAN
(Proceedings of the Lebedev Physics Institute)  211 (1991) 161 (Russian).
\bibitem{4}
S.I.Vinitsky, V.L.Derbov, V.N.Dubovik, B.L.Markovski, and
Yu.P.Stepanovsky,
 Sov. Phys. Usp.  33 (1990) 403.

\bibitem{5}

D.N. Klyshko, Phys. Lett. A140 (1989) 19; ibidem A163 (1992) 349;
R. Bhandari and T. Dasgupta, Phys. Lett., A143 (1990) 170.
\bibitem{6}
R. Tanas and S. Kielich, J. Mod. Opt. 37 (1990) 1935;
R. Tanas and Ts. Gantsog, JINR preprint No E17-91-304 (Dubna, 1991).
\bibitem{7}
V.P.Karassiov,
 J.Sov.Laser Research 12 (1991) 147, 431.
\bibitem{8}
R.Bhandari,
Physica B 175 (1991) 111.
\bibitem{9}
V.P.Karassiov and L.A.Shelepin,
 Teor. i Mat. Fiz.  45 (1980) 54 (in Russian).
\bibitem{10}
R.Delbourgo, J. Phys. A 10 (1977) 1837;
J.R.Klauder and B.-S.Skagerstam, Coherent states.
Applications in physics and mathematical physics (World Scientific,
Singapore, 1985);
A.M.Perelomov, Generalized coherent states and
their applications (Springer, Berlin, 1986).
\bibitem{11}
V.P.Karassiov,
Bulletin Lebedev Phys. Inst. No. 5 (1993) 19.
\bibitem{12}
V.P.Karassiov, in: Proc. 3 Workshop on Squeezed States and
Phys. Lett. A190 (1994) 387; J. Rus. [Sov.]  Laser
Res. 15 (1994) 391; V. P. Karassiov and M. Cervantes, Rev. Mex. Fis.
40 (1994) 227.
\bibitem{13}
M.Berry,
 J. Mod. Opt. 34 (1987) 1401;
T.H.Chyba, L.J.Wang, L.Mandel, and R.Simon,
 Opt. Lett. 13 (1988) 562;
R.Bhandari and J.Samuel,
 Phys. Rev. Lett. 60 (1988) 1211;
R.Bhandari,
 Phys. Lett. A 133 (1988) 1;
M.Kitano and T.Yabuzaki,
 Phys. Rev. A 142 (1989) 321.
\bibitem{14}
S.Pancharatnam,  Proc. Indian Acad. Sci. A 44 (1956) 247;
reprinted in:  Collected Works of S.Pancharatnam (Oxford Univ.
Press, London, 1975).
\bibitem{15}
J.Samuel and R.Bhandari,
 Phys. Rev. Lett. 60 (1988) 2339.
\bibitem{16}
Y.Aharonov and J.Anandan,
 Phys. Rev. Lett. 58 (1987) 1593.
\bibitem{17}
H.P.Yuen,
 Phys. Rev. A 13 (1976) 2226.
\bibitem{30}    
   Agarwal, G.S., Simon, R.,1990, Phys. Rev. A.,{\bf 42, No. 11}, 6924.
\bibitem{31}    
   Hannay,J.,1985, J. Phys. A: Math. Gen.,{\bf V.18}, 221.
\bibitem{32}   
   Berry,M.V., 1985, J. Phys. A: Math. Gen., {\bf 18}, 15.
\bibitem{33}  
   Maamache,M., Provost,J.-P., and Vall\'ee, 1991,
  J. Phys. A: Math. Gen., {\bf 24}, 685.
\bibitem{34} 
   Giavarini,G., Gozzi,E., Rohrlich,D, and Thacker,W.D., 1989,
  Phys. Rev. D., {\bf V.39.} No.10. 3007.
\bibitem{35} 
  J.Christian and A.Shimony. J. Phys. A, {\bf 26}, 5551 (1993).
\bibitem{18} 
C.Aragone, E.Chalbaud, and S.Salamo,
 J. Math. Phys.  17 (1976) 1963.
\bibitem{19}  
 A.Voudras,
 Phys. Rev. A 41 (1990) 1653.
\end{references}
\end{document}